\documentstyle[epsf]{article}
\newcommand{\bee}{\begin{equation}}
\newcommand{\ee}{\end{equation}}
\newcommand{\beea}{\begin{eqnarray}}
\newcommand{\eea}{\end{eqnarray}}
\newcommand{\rme}{{\rm e}}

\begin{document}
\thispagestyle{empty}
\parskip=12pt
\raggedbottom

\def\mytoday#1{{ } \ifcase\month \or
 January\or February\or March\or April\or May\or June\or
 July\or August\or September\or October\or November\or December\fi
%\space\number\day ,
 \space \number\year}
\noindent
\hspace*{9cm} COLO-HEP-369\\
\vspace*{1cm}
\begin{center}
{\LARGE Instantons and Fixed Point Actions in $SU(2)$ Gauge Theory}
\footnote{Work supported in part by
NSF Grant PHY-9023257 
and U.~S. Department of Energy grant DE--AC02--91ER--40672}

\vspace{1cm}

Thomas DeGrand,
Anna Hasenfratz, and Decai Zhu\\
Department of Physics \\
 University of Colorado,
Boulder CO 80309-390

\vspace{1cm}

\mytoday \\ \vspace*{1cm}

\nopagebreak[4]

\begin{abstract}
We describe the properties of instantons in lattice gauge theory 
when the action is a fixed point action of some renormalization group transformation. We present a theoretically consistent
method for measuring topological charge
using an inverse renormalization group transformation. We show that,
using a fixed point action, the action of smooth configurations with
non-zero topological charge is greater than or equal to its continuum value
$8\pi^2/g^2$.
\end{abstract}

\end{center}
\eject

%\end{titlepage}

\section{INTRODUCTION}
The dynamics of
asymptotically free gauge theories are strongly influenced by
 topological effects. In QCD instantons may be responsible for
breaking axial symmetry and resolving the $U(1)$ problem\cite{U1}.
In the large-$N_c$ limit a combination of the masses
 of the $\eta$, $\eta'$, and $K$
mesons is related to the topological susceptibility through
the Witten-Veneziano formula\cite{WV}.

The topological susceptibility $\chi$ is defined as the infinite volume limit 
of
\bee
\chi_t = {{\langle Q^2 \rangle}\over V}
\ee
where $Q$ is the topological charge and $V$ the space time volume.
In QCD $\chi_t$ is a dimension-4 object with no weak coupling expansion.
As such, a calculation of $\chi_t$ in physical units in the continuum
requires nonperturbative techniques.

Over the years there have been many attempts to compute topological 
properties of QCD using lattice Monte Carlo simulations.
A serious problem in lattice studies of topology is the presence of lattice
artifacts. They can arise both from the form of the lattice action
and from the choice of lattice operator to define and measure
topological charge. A lattice action is, in general, not scale invariant, i.e. the
action of a smooth continuum instanton can depend on its size. This leads to 
lattice artifacts, called ``dislocations,''
\cite{DISLOCATION},
that are non-zero charged configurations whose contribution to the
topological charge comes entirely from small localized regions.
If the minimal action of a dislocation is smaller than 6/11 (for SU(2)) 
times the continuum
value of a one-instanton configuration, then dislocations will dominate
the path integral and spoil the scaling of $\chi_t$ \cite{PANDT,BREAKDOWN}.
Difficulties also arise because the topological charge is not conserved on the
lattice. When the size of an instanton becomes small compared to the lattice
spacing it can "fall through" the lattice and its charge disappears. Topological
charge operators can fail to identify this process.

These problems  are circumvented by the use of fixed point 
actions. If one can construct a lattice action and lattice operators
which live on the renormalized trajectory (RT) of some 
renormalization group transformation (RGT), then one's predictions
do not depend on the lattice spacing. A recent series of 
papers\cite{HN,PAPER1,PAPER2,PAPER3} have shown
how to find a fixed point (FP) action for asymptotically free theories,
with explicit examples for spin and gauge models. FP actions share the
scaling properties of the RT (through   one-loop quantum corrections)
and as such may be taken as a first approximation to a RT.
FP actions have scale-invariant instanton solutions with an action value
of exactly $8 \pi^2/g^2$ and, as we will show, one can define a topological 
charge using RG techniques which has no lattice artifacts.

In principle, the questions we are asking and the methods by which 
we answer them are similar to the problems associated with topology
in two-dimensional spin models\cite{INSTANTON,CP3,DEILIA}.
Because we deal with four dimensional gauge theories we face
 additional obstacles in numerical tests due to computer speed and
memory limitations.

In Section 2 we describe the formal properties of instantons
under RG transformations and review the argument that if the action is
a FP action, an instanton solution is scale invariant. We then describe
a consistent
method for measuring topological charge which exploits the invariance
of the instanton solution under RG transformations at the FP. In Section 3
we describe the construction of trial instanton solutions on the lattice
and in Section 4 we review the construction of FP actions and display a
FP action for SU(2) gauge theory for a particular RG transformation.
Finally in Section 5 we show that configurations with nonzero topological
 charge have action greater than or equal to the classical value $8\pi^2/g^2$
while configurations with action less than the classical value have zero topological charge. In this paper we do not address the calculation of the
topological susceptibility via Monte Carlo simulation.

\section{Formal Considerations for Ideal Instantons}

We consider an SU(N) pure gauge theory\footnote{The actual numerical
analysis and simulations in this paper were done for SU(2). The
equations are written for general N if not indicated otherwise.} on
the lattice and the RG transformation
\bee
\rme^{-\beta' S'(V)} = \int DU
\exp\left\{ -\beta \left( S(U)+T(U,V)\right) \right\}  ,        \label{RGT}
\ee
where $U$ is the original link variable, $V$ is the blocked link
variable and $T(U,V)$ is the blocking kernel that defines the
transformation. At $\beta = \infty$ the transformation becomes
a steepest descent relation with a fixed point solution
\bee
S^{FP}(V)=\min_{ \{U\} } \left( S^{FP}(U) +T(U,V)\right),  \label{STEEP}
\ee
where
\bee
T(U,V)=-{\kappa\over N}
\sum_{n_B,\mu} \left[ {\rm Re Tr} \left( V_{\mu}(n_B)
Q_{\mu}^{\dagger}(n_B)\right) -
\max_{W} \left\{ {\rm Re Tr} (W Q^{\dagger}) \right\} \right].
                                              \label{TEQN}
\ee
In Eqn.~(\ref{TEQN}) $W \in {\rm SU}(N)$ and the $N \times N$ complex
matrix $Q_{\mu}(n_B)$ is the block average. The 
block transformation which we consider here  is a scale-two
Swendsen type \cite{RNG} transformation,
that is referred to as ``type I'' RGT  and 
$Q$ is defined in Fig.~5 in \cite{PAPER1}.
For further details we refer the reader to  Ref. \cite{PAPER1}.
Eqn.~\ref{STEEP} transforms a 
 coarse lattice  $\{V\}$ that has structure on the scale
of one coarse lattice spacing to a fine lattice $\{U\}$ with half the lattice
spacing of $\{V\}$. 
We will refer to  this procedure as
``inverse blocking.''

In the continuum the action is conventionally defined as
\bee
{\cal S} =  -{1\over 4}\int d^4x F_{\mu \nu} F^{\mu \nu}
\ee
and the continuum instanton action is $ {\cal S}=8\pi^2/g^2$. 
In lattice conventions ${\cal S} = \beta S$
with $\beta= 2N/g^2$ for $SU(N)$, and the equivalent result for
the instanton action is $S_I = 4\pi^2/N$. In all tables and figures
in the paper which deal with actions of instanton-like configurations
we will present results in units of $S_I$ so that the reader will
not be forced to deal with the particular conventions we use for $S$
(such as sums over space-time indices, etc.).
In order to avoid domination of
the functional integral by dislocations, the action of a configuration
with nonzero topological charge must satisfy the constraint
\bee
S > {{48 \pi^2} \over {11N^2}},
\ee
or $S/S_I > 6/11$ for $SU(2)$.

The  continuum instanton action  is scale invariant. 
In general this is not the case on the lattice, as is well known, for
example, for the plaquette Wilson action. On the other hand FP
actions are scale invariant.
The proof of that statement was given in Ref. \cite{HN}
for the $d=2$ $\sigma$--model and easily generalizes for asymptotically free
gauge models.
If $\{ V \}$ is a solution of the classical equations of motion, so that
$\delta S^{FP}/\delta V =0$, then 
\bee
{{\delta S^{FP}}\over{\delta V }}={{\delta T}\over{\delta V }}
+ {{\delta U}\over{\delta V }}{{\delta}\over{\delta U }}(S^{FP}+T)
 = 0.\label{STEEPER}
\ee
Eqn.~\ref{STEEP} implies that 
for the minimizing configuration $\{ U(V)\}$ the
second term in Eqn.~\ref{STEEPER} vanishes.
Then it follows that $\delta T/\delta V =0$, too, or that $T(U,V)$ 
also takes its minimum value,  zero. Thus
\bee
\left. {\delta S^{FP}(U) \over \delta U }
\right|_{ U=U(V) } =0,                                      \label{11}
\ee
and $S^{FP}(V)=S^{FP}(U(V))$.
According to this result the FP action has exact scale invariant
instanton solutions with action equal to
$8\pi^2/g^2$ as in the continuum theory.

It also follows that the inverse blocking transformation 
does not change the topological charge of a
configuration.
According to Eqn.~\ref{STEEP} the action on the fine lattice
is always equal to or smaller than the action on the coarse lattice,
because the blocking kernel $T(U,V)$ is positive-definite.
Thus a coarse configuration that has zero topological charge
will inverse block into a configuration with lower action, and after
many levels of inverse blocking the action on the finest lattice will
go to zero. In contrast,
the action of a continuum instanton configuration will not change
under blocking.
A coarse configuration consisting of an instanton
plus fluctuations, whose action is greater than $S_I$, will
inverse block to a configuration whose action is closer to $S_I$,
and under repeated inverse blocking will approach $S_I$ arbitrarily closely.
The size of the instanton grows by a factor of two under the inverse
blocking, so the physical picture of the many-times inverse blocked
lattice configuration is a large smooth instanton with its appropriate action.

There are two types of definitions of topological charge
 in the
literature.
One is the
``algebraic'' or ``field theoretic'' definition\cite{ALGEBRA} where
some lattice discretization of the continuum operator $F \tilde F$
is measured. 
The geometric definition\cite{GEOMETRIC} reconstructs a fiber bundle from the
lattice gauge field and identifies the second Chern number of this bundle with
the topological charge.

Both definitions are correct and produce equivalent results
on smooth  configurations.
On a coarse lattice with a sufficiently rough gauge
configuration  both definitions of the topological charge break down.
The field theoretical definition  does not yield integer values of the charge
and to obtain a continuum charge the lattice charge is multiplicatively
renormalized by a coupling-dependent  (and action-dependent)
factor $Z(\beta)$.
The algebraic definition could be improved by using a
FP instanton charge as described by Ref. \cite{DEILIA}
--if it could be done nonperturbatively.
We elect not to pursue the algebraic method in this paper due to
 the complication of the $Z(\beta)$ factor.
The geometric definition also fails as it often identifies the remnants of an
instanton that has fallen through the lattice as a non-zero charge object thus
predicting an artificially large topological susceptibility.

Eqn.~\ref{STEEP}  provides a way  to smooth rough configurations in a way
which preserves their topological properties.
We define the topological charge of a configuration by first inverse blocking
it to a sufficiently smooth configuration and then measuring the charge 
 on  the fine
lattice. 
If the initial configuration is sufficiently rough, one might need to inverse
block more than once to get a sufficiently smooth configuration
to reliably measure $Q$.
Since  the geometric definition works on rougher configurations than the field
theoretic one, we elected to use the geometric method to measure $Q$.
Practical
considerations of computer memory and speed restrict us to a single
step of inverse blocking.

Notice that the above  definition of topological charge requires the 
specification of some renormalization group transformation and its
associated FP action. Inverse blocking with an arbitrary action is
not guaranteed to preserve topological charge from one level of 
blocking to the next.

Either the algebraic or geometric definition of topological charge
could be--and has been--combined with some other 
 smoothing algorithm. One popular
smoothing algorithm is ``cooling.'' The idea behind cooling is that one
takes a gauge configuration and performs some local minimization of the
action; this minimization is supposed to eliminate
 short-distance fluctuations while preserving long-distance structure.
Unfortunately, cooling is not a trustworthy indicator of topological
structure for rough gauge configurations. The topological
charge of a cooled configuration depends on the particular action whose
value is to be minimized \cite{VANBAALETC}, and if the action is not
constrained to be $S_I$ for all $Q\ne0$ configurations, the act
of cooling could change $Q$.
As a practical problem, the particular algorithm used to cool may
be so efficient that it does not find local minima of its action but instead
global minima: the lowest global minimum is some gauge transform of
the identity.
While it may be possible to find a cooling algorithm which reproduces
the results of inverse blocking, we elect in this work to use only
the theoretically reliable inverse blocking method for measuring topological
charge.

\section{Practical Considerations for Ideal Instantons}

We want to demonstrate that an FP action is indeed scale invariant by
showing that the action of an instanton that is the solution of the equations of
motion is independent of its size.
We will parameterize these solutions by a radius or scale
factor $\rho$ and measure the topological charge $Q(\rho)$ and
the action $S(\rho)$. We want to show that when $Q=1$ $S(\rho)\ge S_I$
and when $S(\rho)\le S_I$, $Q =0$.
We will also illustrate that our definition of the topological charge is free of
dislocations, at least for smooth instantons.

We must first deal with a couple of technical problems.

\subsection{Finite volume effects}
We preface this section by writing down a few formulas for the
gauge potential for a single continuum instanton:
\bee
A_\mu(x) = -i f(x)  g \partial_\mu g^\dagger        \label{INST}
\ee
where $x$ is a Euclidean four-vector, 
\bee
f(x) = {{x^2}\over{x^2+\rho^2}}        \label{13}
\ee
is the shape factor, and
\bee
g = {1 \over x } (x_0 + i \vec x \cdot \vec \sigma) .      \label{GT}
\ee

On a periodic lattice a  single instanton is not a solution of the classical
equations of motion; it does not obey periodicity.
If we go ahead and lay down an instanton-like solution
on a periodic torus of size $L$, we find that its action diverges
linearly:
\bee
S(L)= S_0 + O(L)    .                                  \label{12}
\ee
due to the discontinuity of the field configurations
at $x_0 = \pm L/2$.

To ameliorate this problem we follow Pugh and Teper\cite{PANDT}
and consider instead trial solutions made of an instanton and a 
superimposed dislocation. (The dislocation can be regarded as the
remnant of a small anti-instanton that fell through the lattice.)
Take the solution defined by Eqn. \ref{INST} and perform a singular gauge
transformation on it
\bee
V_\mu(x) = g^\dagger(x)  U_\mu(x) g(x+\hat\mu)             \label{IPLUSDEF}
\ee
where $g$ is defined in Eqn. \ref{GT}.
It is easy to show that on a periodic lattice
the finite-volume correction to the action
of this configuration is
\bee
S(L)= S_0 + O(1/L^3)                                      \label{14}
\ee
As we study discretized instantons, we will do computations
 on many different lattice sizes $L$
and extrapolate results for the action to infinite volume using
\bee
S(L)= S_0 + a_1/L^3 + a_2/L^5 .                      \label{EXTRAP}
\ee

\subsection{Short distance effects}
The discretization of the instanton solution on the lattice
will also show lattice artifacts when the size of the instanton solution
(parameterized by $\rho$) is of the order of the lattice spacing.
A true instanton solution of a FP action has no such artifacts.

To place an instanton-like solution on a lattice we
begin with a vector potential $A_\mu(x)$ defined via Eqn. \ref{INST}, with
$x$ measured at any continuum point with respect to an origin $x_0$.
Define link variables on a lattice on sites labeled by an integer $n$
by approximately constructing the path-ordered exponential between
neighboring sites.
That is, define
\bee
U_\mu(na) = \prod_j U_\mu(x=na+j\Delta x\hat\mu)
\ee
where
\bee
U_\mu(x=na+j\Delta x\hat\mu)= \exp(i\Delta x A_\mu(x=na+j\Delta x\hat\mu))
 \label{UUU}
\ee
and
$A_\mu(x)$ is given by Eqn. \ref{INST}.
We typically broke the lattice spacing up into 20 intervals
($\Delta x = a/20$).
The interpolation corresponds to smoothing the classical solution along the
lines of the lattice.
Finally, we perform the singular gauge transformation of Eqn. \ref{IPLUSDEF}
on the link variable of Eqn.~\ref{UUU}.
So far this definition follows Pugh and Teper\cite{PANDT}.

This procedure  does not guarantee that we have constructed
a solution for the equations of motion of our FP action. Short distance
roughness in the solution, especially for small instantons, is still present.
In  order to remove this roughness we block the instanton  configuration
(perhaps repeatedly) using the RGT of Eqn. \ref{RGT}.
Blocking  at $\beta\rightarrow\infty$ moves the configuration
towards the FP and transforms the instanton to a solution of the equations of
motion.
Blocking does not preserve the topological charge. With each blocking step the
size of the instanton is halved and it might disappear from the lattice. Both the
value of the action and the topological charge operator should signal when that
happens.

We consider trial solutions made after one level of blocking
 ($40^4 \rightarrow
20^4$, $32^4 \rightarrow 16^4$,  $20^4 \rightarrow
10^4$  and $16^4 \rightarrow 8^4$) and after
 two levels of blocking ($40^4 \rightarrow 10^4$, $32^4 \rightarrow 8^4$, $16^4
\rightarrow 4^4$).
In addition we consider two kinds of instanton-like solutions,
 which we call $c_1$ and $c_2$ instantons, whose centers
are located at $x_0 = (L/2+1/2,L/2+1/2,L/2+1/2,L/2+1/2)$ and
$x_0 = (L/2,L/2,L/2+1/2,L/2+1/2)$  respectively  on an $L^4$ lattice.

\section{Construction of a FP action}

 Determining the properties of instantons using a FP action requires
knowing the action.
The construction of a FP action quadratic in the gauge fields,
valid for smooth gauge configurations, was described in \cite{PAPER1}.
However, instanton solutions with small $\rho$ are not smooth
and we need a parametrization of the FP action valid on rough configurations as
well.

Our parametrization of the action is based on powers of the traces of the
loop products $V_{\it C} =  \Pi_{\it C} V_\mu(n)$ (for SU(N))
where $\it C$ is an arbitrary closed path
\bee
S^{FP}(V) = {1 \over N} \sum_{\it C} ( c_1({\it C})(N-
{\rm Re}{\rm Tr}(V_{\it C})) +
                       c_2({\it C})(N-{\rm Re}{\rm Tr}{\rm Tr}(V_{\it C}))^2+ ...
\label{ACTIONPAR}
\ee

Assume we know the value of the fixed point action on a set of gauge
configurations $\{ V\}$. As we can measure the traces of the loop products in Eqn.
\ref{ACTIONPAR}, we obtain a set of linear equations, one for each $\{ V\}$
configuration, for the coefficients $c_i({\it C})$. Consequently the parameters of
the fixed point
action on a given set of $\{ V\}$ configurations can be obtained from a
linear fit.

\begin{figure}[htb]
\begin{center}
\vskip 10mm
\leavevmode
\epsfxsize=90mm
\epsfbox{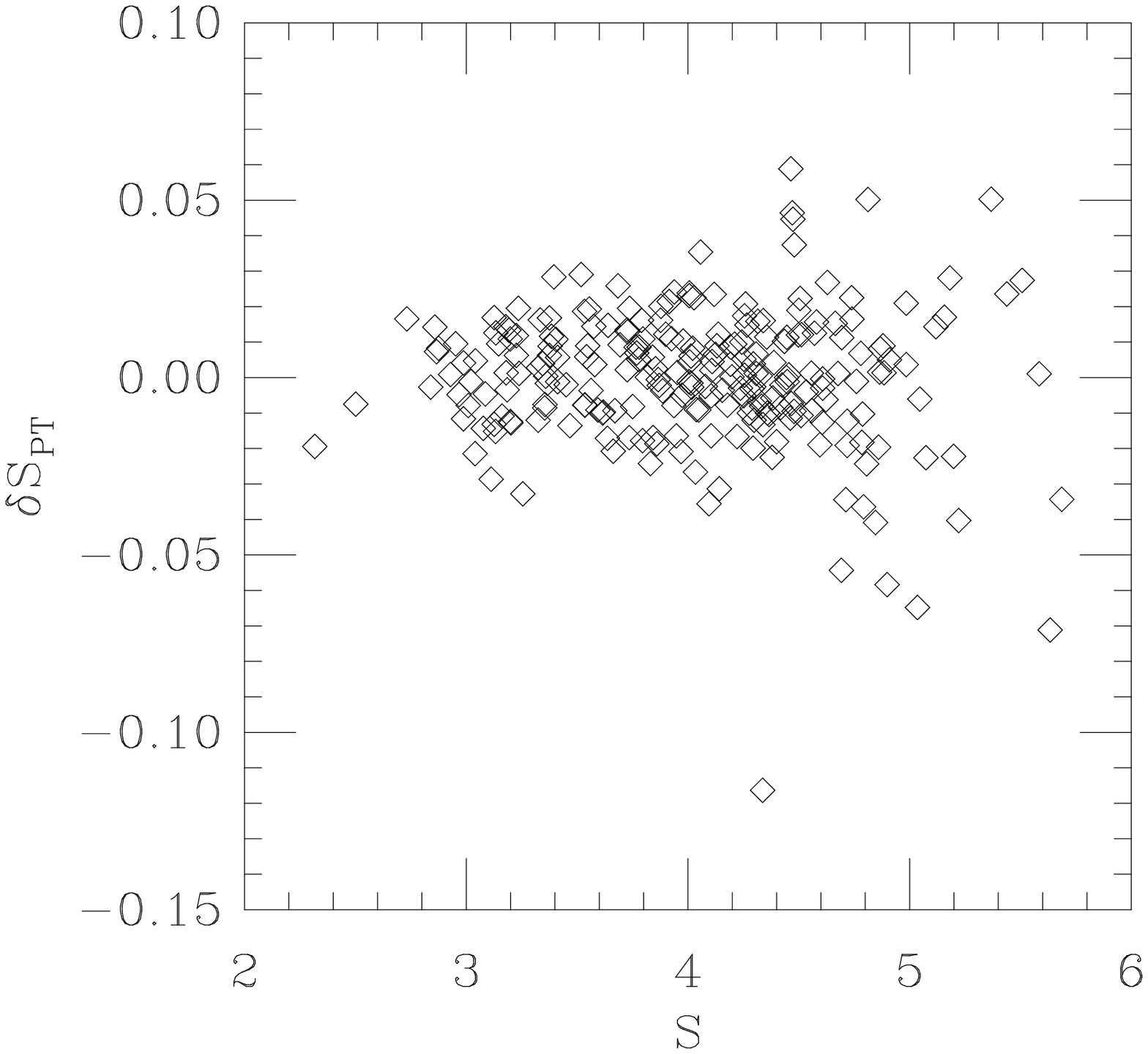}
\vskip 10mm
\end{center}
\caption{ The perturbative corrections
$\delta S_{PT}=S^{FP}(U_0)-S_0(U_0)$ for a set of configurations
using the type I RGT, for the quadratic limit of the FP action.}
\label{fig:delta_pt_vs_S_su2}
% FIGURE_FILE delta_pt_vs_S_su2.ax
\end{figure}

All that is left is to calculate the value of the FP action on a given
configuration. The inverse blocking procedure of Eqn.~\ref{STEEP} can be used for
that. For any coarse configuration $\{ V\}$ we can find a fine configuration
$\{U^{FP}\}$ that minimizes the right hand side of Eqn.~\ref{STEEP} and calculate
the FP action as
\bee
S^{FP}(V) = S^{FP}(U^{FP})+T(U^{FP},V).
\ee
For this procedure we need the FP action on the fine configurations only. As the
$\{U\}$ configurations are very smooth, the FP action is well approximated by its
analytically  derived quadratic form $S_q^{FP}$ on the fine configurations.
$S_q^{FP}$ is still too complicated for the numerical minimization. Fortunately  
it is possible to replace $S_q^{FP}$ with a simpler action
$S_0$ in the minimization if its corresponding fine configuration $\{U_0\}$ is
close to $\{U^{FP}\}$ and the difference $S_q^{FP}-S_0$ can be corrected
perturbatively. If
\bee
S'(V) = min_{U}(S_0(U)+T(U,V))=S_0(U_0)+T(U_0,V),
\ee
then
\bee
S^{FP}(V)= S_q^{FP}(U_0)+T(U_0,V) = S^{FP}(U_0)+S'(V)-S_0(U_0) ,
\ee
assuming that $S_q^{FP}(U_0)-S_0(U_0)$ is small. We took for $S_0$ the
 single plaquette action
\beea
S_0(U) & = &0.9113(2-{\rm Tr} U_P) + 0.4564(2-{\rm Tr} U_P)^2
- 0.7723(2-{\rm Tr} U_P)^3 \nonumber \\
   &  & + 0.3007(2-{\rm Tr} U_P)^4 \nonumber \\
\eea
and found that the perturbative corrections are about 0.5\% as illustrated in 
Fig. \ref{fig:delta_pt_vs_S_su2} for a set of
sample $\{V\}$ configurations.

FP actions, even the most local ones, can be fairly complicated. At very large
correlation lengths the FP action is well described by its quadratic form,
given by the $c_1( {\it C})$ coefficients of Eqn.~\ref{ACTIONPAR}.
 As the
correlation length decreases the higher representation terms become important and
it is necessary to determine the $c_i( {\it C}), i>1$ coefficients as well.
At even smaller correlation lengths it is possible that new type of operators,
 which
were not present in the quadratic limit 
 (products of disconnected loops, for example),
will become important.

We parametrized the fixed point action with the 12 loops present in the
quadratic limit, each in 4 different representations, i.e. 
 using the first 4 powers in the series of Eqn.~\ref{ACTIONPAR}. We
found that a 48 parameter fit describes the fixed point action well if the
lattice spacing of the coarse configurations is $a \le 1/3T_c$.

The couplings of a 48 parameter fit are given in 
Table \ref{tab:couplingsI}. In this fit we used
about 500  configurations
generated with the Wilson action,  all with $a\ge1/3T_c$, 
and about 20 smooth instanton
configurations with $\rho>1.0$
that we created with the procedure described in Sect. 3. The FP
action  value of the instanton configurations were fixed by requiring that on infinite
volume their action extrapolates to the theoretical value $S=S_I$.

Now  we turn our attention to
the scale invariance of the full FP action.

\begin{table*}[hbt]
% space before first and after last column: 1.5pc
% space between columns: 3.0pc (twice the above)
\setlength{\tabcolsep}{1.5pc}
\caption{Couplings of the Type I RGT   fixed point action 
given in terms of
loops defined in Table 3 of Ref. 7. The quadratic couplings are labeled by $c_1$,
and the coefficients of higher powers of $(2- {\rm Tr}U(C))$
from the fit described in Sec. 4 are also shown.}
\label{tab:couplingsI}
%\begin{tabular}{l|cccccc}
\begin{tabular*}{\textwidth}{@{}l@{\extracolsep{\fill}}lcccccc}
\hline
loop & $c_1$ & $c_2$ & $c_3$  &$c_4$ \\
    1 &   .6744  &-.00716  & .0934  & -.0190 \\
    2 &  -.02 & -.170& .117 & -.0202 \\
    3 &  .012  & -.136  & .0864 & -.0141 \\
    4 &  .005 & -.101  &  .0729 & -.0112 \\
    5 &  -.0031 & .0600 & -.0380 & .00708 \\
    8 &  -.0035 & .0281& -.0252 & .00426 \\
    9 &  .0027 & -.112 & .0697& -.0103 \\
    15 &  -.0024 & .0542 & -.0346 & .00531 \\
    16 &  .0013 & .0348  &-.0242 & .00416 \\
    20 &  .003 & .106 & -.0695 & .0122 \\
    23 &  .0032 & -.0550 & .0385 & -.0060 \\
    25  & .0024 & -.0343 & .0077 & .000053 \\
\hline
\end{tabular*}
\end{table*}

\section{The action and topological charge of instanton configurations}
 
We measure both the topological charge and the value of the fixed point action
using the inverse block transformation. 
For each trial configuration $\{ V \}$ we
create a fine configuration $\{ U \}$ according to  Eqn.~\ref{STEEP}. The fine
configurations are sufficiently smooth that the topological charge can be measured
by the geometric method
\bee
Q(V) = Q_{geom}(U),
\ee
and the minimization gives the value of the fixed point action as well
\bee
S^{FP}(V) = S^{FP}(U)+T(U,V) \label{SINVB}.
\ee
 Measuring the action using Eqn. \ref{SINVB} rather than by direct
evaluation on the coarse lattice exploits the fact that we know the
FP action better on the fine configuration since 
the gauge field is smoother there.
For the minimization procedure we need the FP action on the fine configurations
only. 
 
We found that for the instanton configurations a Symanzik type action
\bee
S_0= {5 \over 3} (N-TrU_{1\times1}) - {1 \over 12} (N- TrU_{1 \times 2})
\ee
was efficient in the minimization.
The difference $S^{FP}(U_0)-S_0(U_0)$ was less then a percent with this choice.
 
We have four approximations to a smooth instanton solution:
$c_1$ or $c_2$ origins, single or double blocked.
When the four approximations do not give identical results for
the action or topological charge, we know that the configurations do not satisfy
the equations of motion. They could be instanton configurations with small
scale
fluctuations or configurations where the instanton fell through the lattice. In
the former case we expect $Q=1$ and $S>S_I$, in the latter $Q=0$ with arbitrary
action.

Table  \ref{tab:summary}
 gives the value of $S_0(U_0)$,  $T(U_0,V)$, 
$S_{fine}=S^{FP}(U_0)$,
$S^{FP}(V)=S^{FP}(U_0)+T(U_0,V)$, $Q_{geom}(U_0)$ and $Q_{geom}(V)$ 
for a series of $c_1$
configurations blocked once from $16^4$ to $8^4$ lattices.
The instanton radius is measured on the coarse ($8^4$) lattice.
We inverse block the $8^4$ configuration to $16^4$ and measure its action
and charge there.
At large $\rho$ values the topological charge is $Q=1$ and $T(U_0,V)$ is 
small, as one
would expect for a smooth instanton. The action $S^{FP}(V)$ is slightly larger
than $S_I$ due to the boundary effects. As  $\rho$ drops below 0.88, $T$
increases about 2 orders of magnitude and $S^{FP}(V)$ falls below $S_I$. The
instanton has fallen through the lattice. The topological charge measured
 on the fine
configuration correctly describes the situation. The charge measured on the coarse
lattice is still $Q(V)_{geom}=1$. The geometric
definition incorrectly identifies a dislocation with a
$Q=1$ instanton.

\begin{table*}[hbt]
\caption{Action and charge of a set of instanton configurations
blocked from $16^4$ to $8^4$ lattices.}
\label{tab:summary}
%\begin{tabular}{l|cccccc}
\begin{tabular*}{\textwidth}{@{}l@{\extracolsep{\fill}}lcccccc}
\hline
$\rho$ &  $T/S_I$ & $S_{fine}/S_I$ & $(T+S_{fine})/S_I$ &
 $Q_{geom}(U)$  & $Q_{geom}(V)$ \\
 \hline
0.5 &  0.119102 & 0.194042 & 0.313144  & 0 & 0  \\
0.8 &  0.153455 & 0.698940 & 0.852395 & 0 & 1  \\
0.85 &  0.145686 & 0.789170 & 0.934856 & 0 & 1  \\
0.86 &  0.138044 & 0.816921 & 0.954965 & 0 & 1  \\
0.88 &  0.001922 & 1.004702 & 1.00662 & 1 & 1  \\
0.89 &  0.001359 & 1.004853 & 1.00621 & 1 & 1  \\
0.9 &  0.001006 & 1.005316 & 1.00632 & 1 & 1  \\
0.91 &  0.001010 & 1.005891 & 1.0069 & 1 & 1  \\
0.92 &  0.000895 & 1.006427 & 1.00732 & 1 & 1  \\
0.95 &  0.000838 & 1.007948 & 1.00879 & 1 & 1  \\
1.0  & 0.000909 & 1.010088 & 1.011 & 1& 1 \\
1.2 &  0.001611 & 1.017157 & 1.01877 & 1 & 1  \\
1.3 &  0.002159 & 1.021950 & 1.02411 & 1 & 1  \\
1.5 &  0.003644 & 1.036813 & 1.04046 & 1 & 1  \\
\hline
\end{tabular*}
\end{table*}

\begin{figure}[htb]
\begin{center}
\vskip 10mm
\leavevmode
\epsfxsize=90mm
\epsfbox{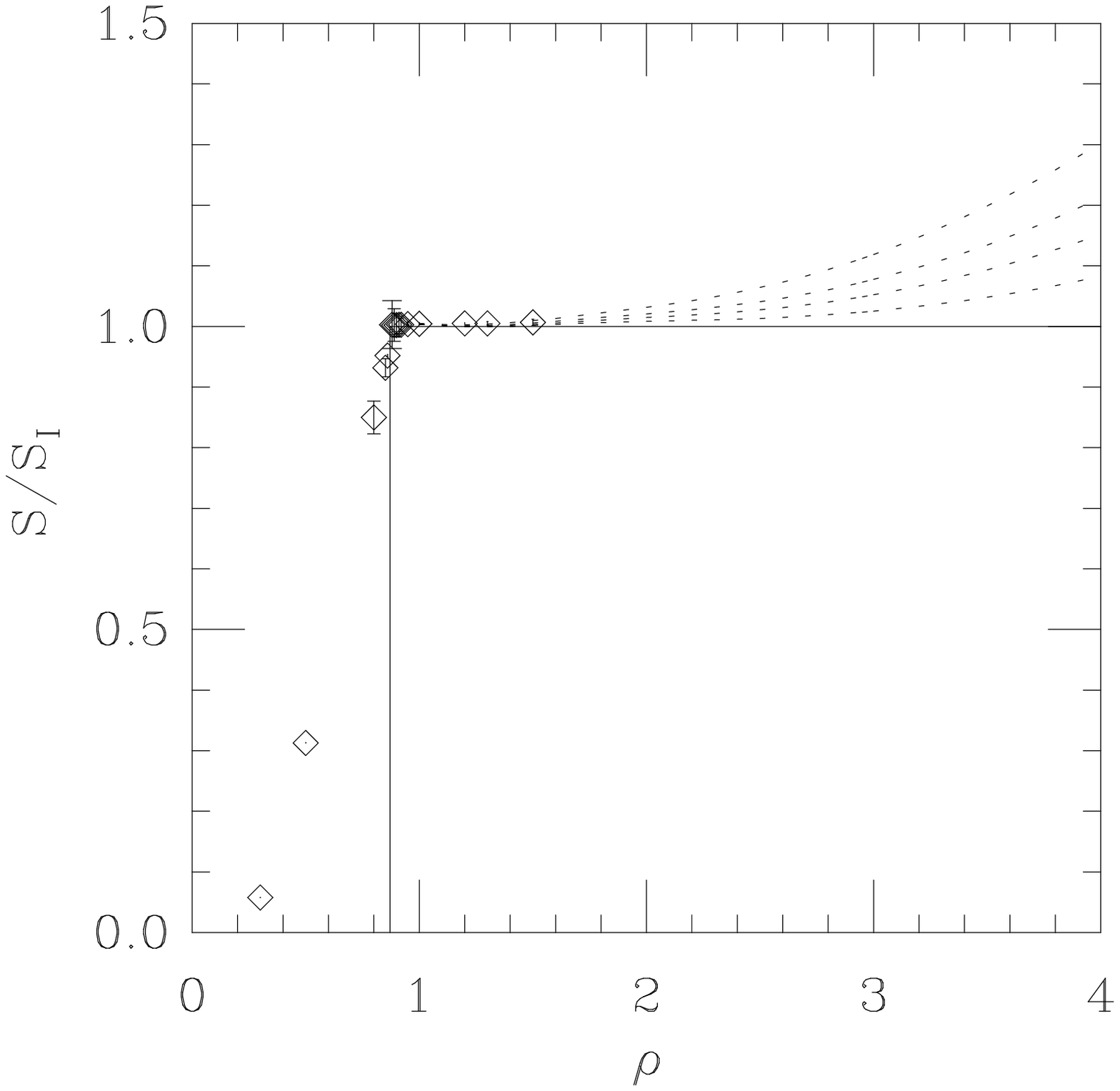}
\vskip 10mm
\end{center}
\caption{ Action of once blocked $c_1$
instanton configurations  computed using Eqn. 3.
The error bar represents the uncertainty in the action due to choice
of different 48 parameter actions and reflects our inability to
completely reconstruct the FP action.
The vertical line corresponds to the $Q=0  \rightarrow Q=1$ boundary.
}
\label{fig:inst_action_vs_rho_b1_c1}
% FIGURE_FILE inst_action_vs_rho_b1_c1.ax
\end{figure}

We cannot specify the FP action or the action of an instanton candidate
to arbitrary accuracy.
There are several sources for the uncertainty of the action. For larger $\rho$ the
infinite volume extrapolation introduces error. For smaller instantons the main
source of uncertainty is the specific parametrization of the FP action.
Table \ref{tab:couplingsI} 
gives a 48 parameter form obtained from a fit to a data set which
included both instantons and a collection of non-instanton
``typical gauge field'' configurations   with $a\ge1/3T_c$.
The small core instantons are rougher than these
configurations. If we use a different set of configurations with the same
qualitative properties we obtain a  slightly different parametrization. The new
parametrization gives the same action for the configurations with $a>1/3T_c$ but
up to 10\% different value for the small radius instantons
(using Eqn. \ref{SINVB}). We illustrate those differences 
by presenting a profile of action vs.~$\rho$
for a set of once-blocked $c_1$ instantons in 
Fig. \ref{fig:inst_action_vs_rho_b1_c1}. The uncertainty in the action is
represented as an error bar on the value of the action, obtained by taking
two equally good parameterizations of the FP action and computing the
action of the test configuration using Eqn. \ref{SINVB}.
The dotted lines show the extrapolation to infinite volume using
Eqn. \ref{EXTRAP} and the couplings of Table \ref{tab:couplingsI}.
The value of the topological charge 
 (measured on the inverse blocked configuration) is overlaid on the figure.
Notice that the action is consistent with the continuum value $S_I$
while $Q=1$ and falls below that number when $Q=0$. Notice also that
while the difference in the action is larger for configurations which
do not carry topological charge, the two parameterizations agree closely
for the value of the action for an instanton configuration.
 
\begin{figure}[htb]
\begin{center}
\vskip 10mm
\leavevmode
\epsfxsize=90mm
\epsfbox{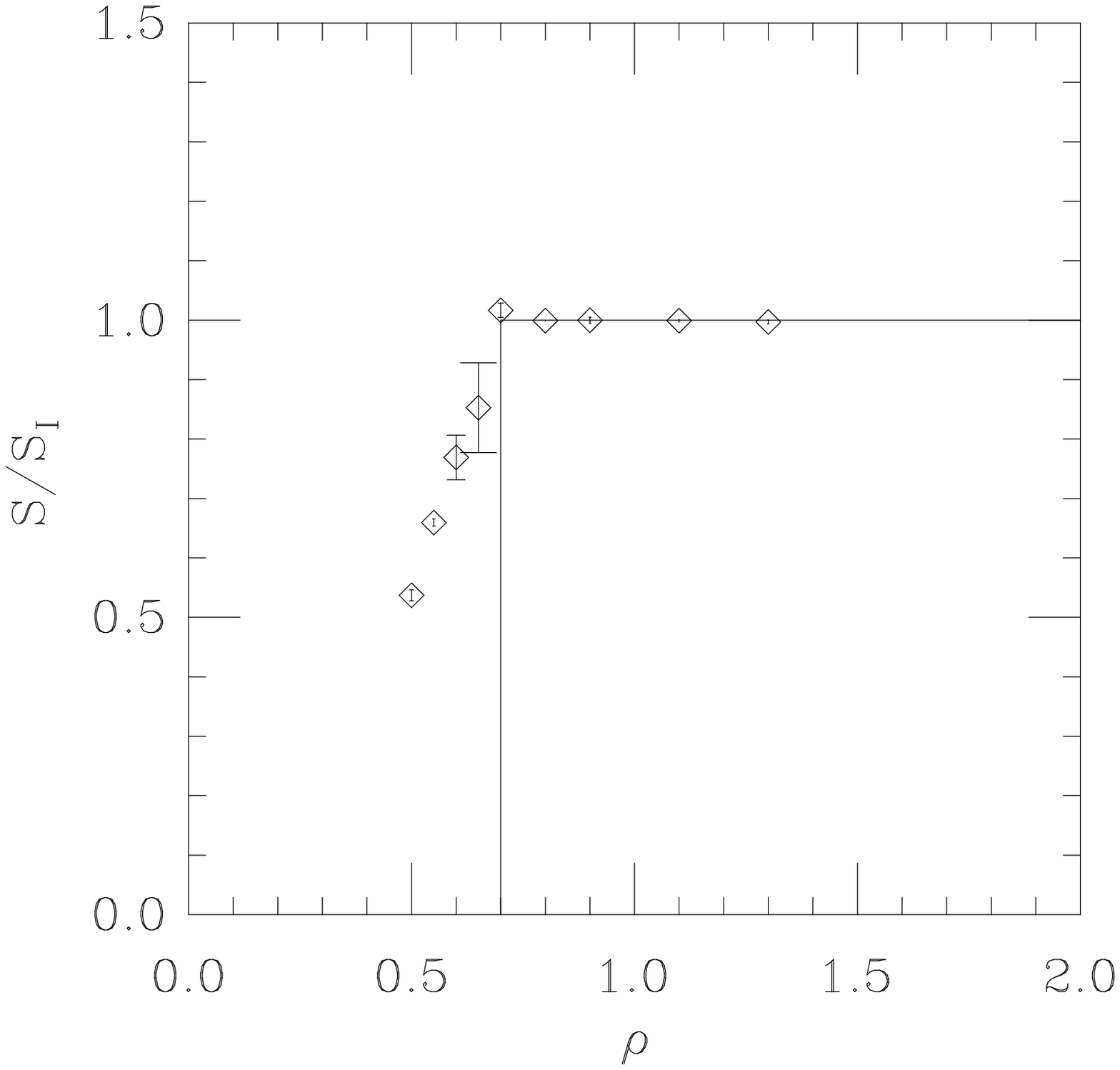}
\vskip 10mm
\end{center}
\caption{ Action of once blocked $c_2$
instanton configurations presented as in Fig. 2.}
\label{fig:inst_action_vs_rho_b1_c2}
% FIGURE_FILE inst_action_vs_rho_b1_c2.ax
\end{figure}

\begin{figure}[htb]
\begin{center}
\vskip 10mm
\leavevmode
\epsfxsize=90mm
\epsfbox{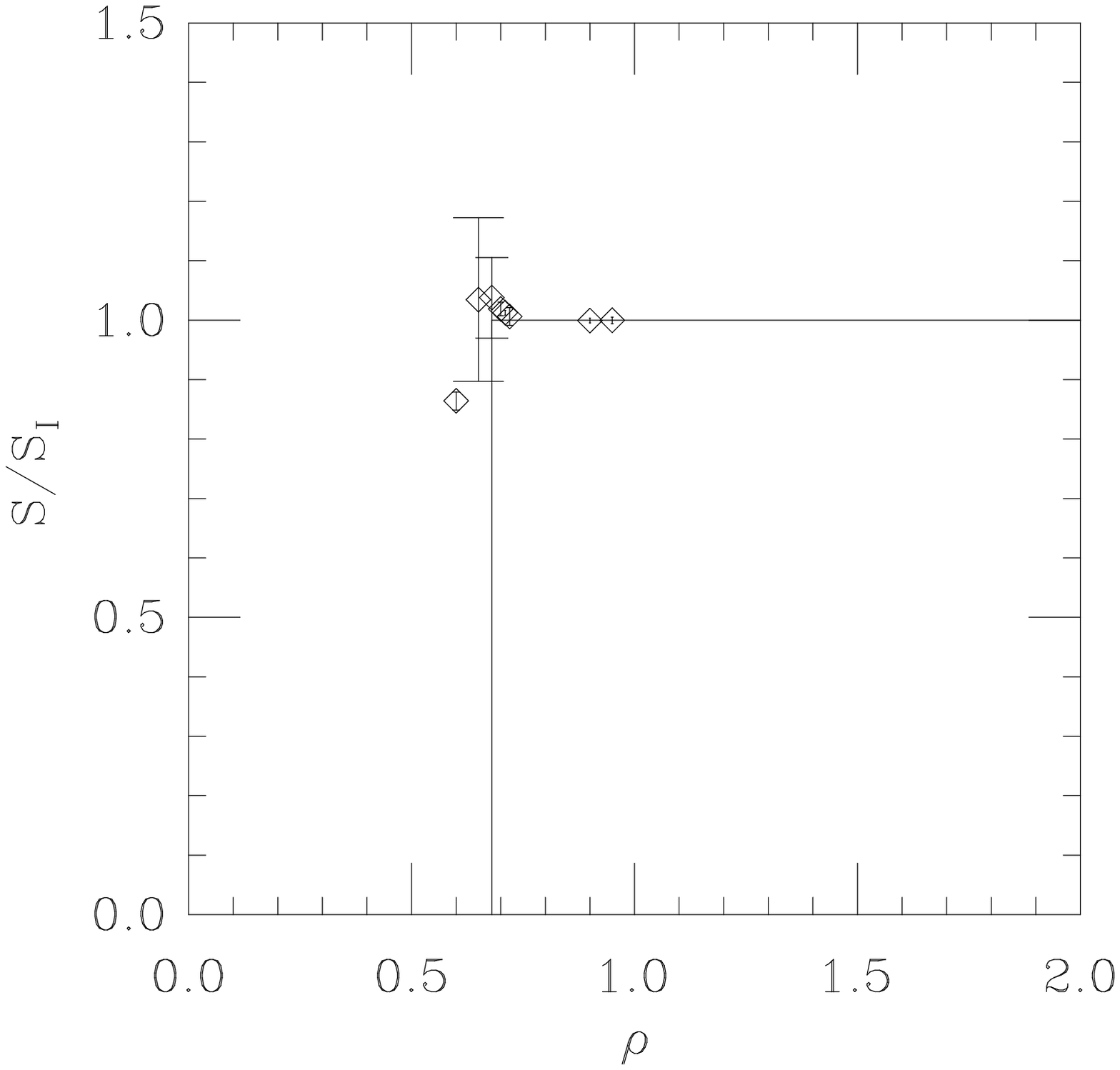}
\vskip 10mm
\end{center}
\caption{ Action of twice blocked $c_1$
instanton configurations presented as in Fig. 2.}
\label{fig:inst_action_vs_rho_b2_c1}
% FIGURE_FILE inst_action_vs_rho_b2_c1.ax
\end{figure}

The corresponding plots for once blocked $c_2$ and twice blocked $c_1$ instantons 
show the same qualitative features 
(Figs. \ref{fig:inst_action_vs_rho_b1_c2} and \ref{fig:inst_action_vs_rho_b2_c1})
 but the quantitative differences
indicate that one or two blocking steps do not get rid of all the short distance
fluctuations of small radius instantons.

\section{Instantons and the Wilson Action}

Our theoretically consistent definition of the topological charge is 
closely related
to a FP action. In principle the same configuration can have different
topological charge and action depending on the FP action we use to measure it,
though in practice that happens only for small core instantons. The Wilson
plaquette action is not a FP action of any RG transformation. It is not
possible to define the topological charge of a configuration with respect to the
Wilson action.

Most of the topological calculations so far used the Wilson action
 and the charge
was measured directly either on the original lattice or on its cooled version.
The recent overimproved cooling technique relies on actions improved according to
their properties on smooth instanton configurations,
whose topological charge is identified by the direct
geometric definition \cite{VANBAALETC,DEFOR}.

\begin{figure}[htb]
\begin{center}
\vskip 10mm
\leavevmode
\epsfxsize=90mm
\epsfbox{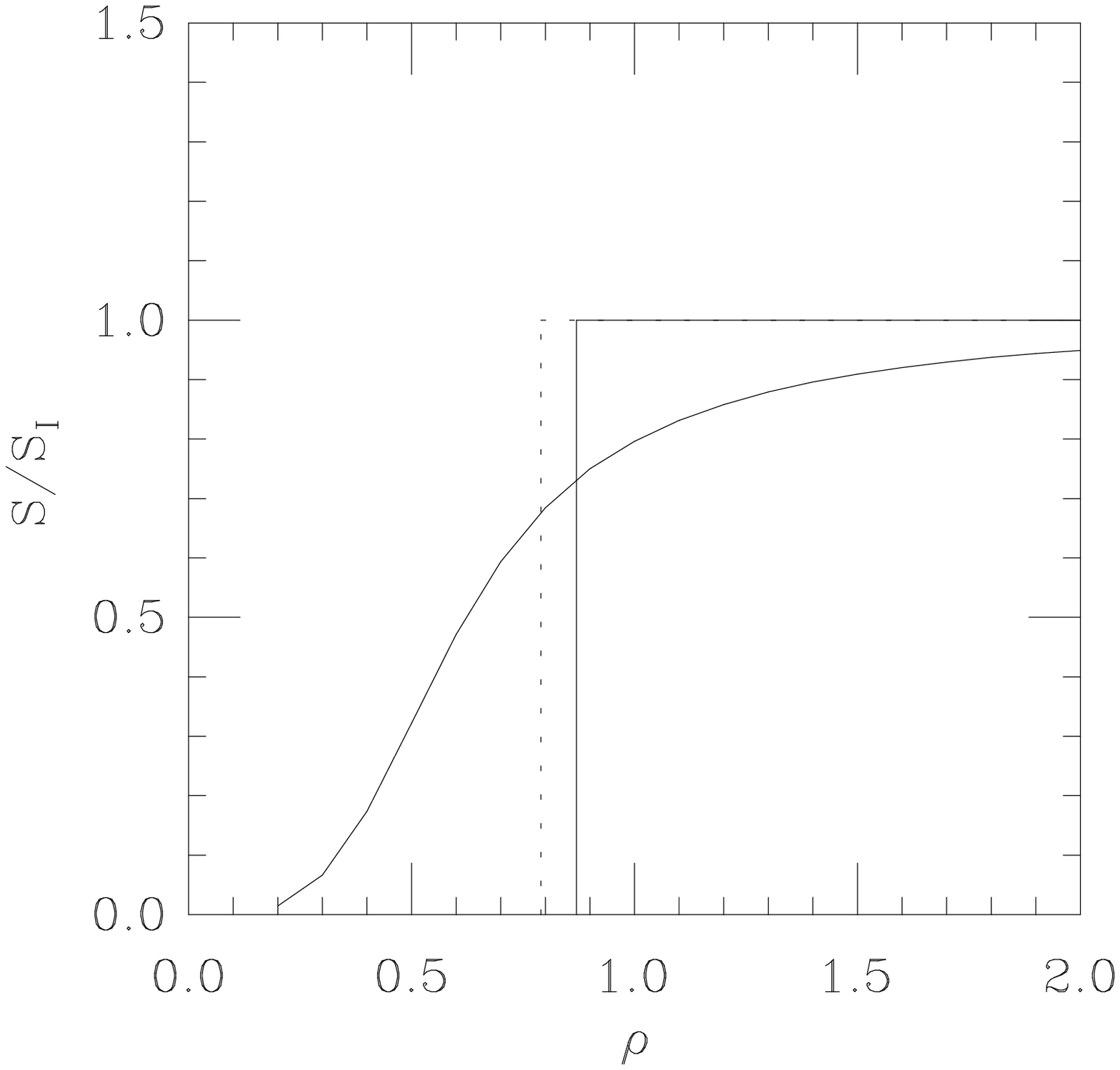}
\vskip 10mm
\end{center}
\caption{ Action of single blocked $c_1$
instanton configurations computed using the Wilson action.
The dotted line indicates the
$Q=0 \to Q=1$ boundary on the original coarse lattice while the solid line is the
boundary on the inverse blocked configuration.}
\label{fig:sbc1w}
% FIGURE_FILE sbc1w.ax
\end{figure}

\begin{figure}[htb]
\begin{center}
\vskip 10mm
\leavevmode
\epsfxsize=90mm
\epsfbox{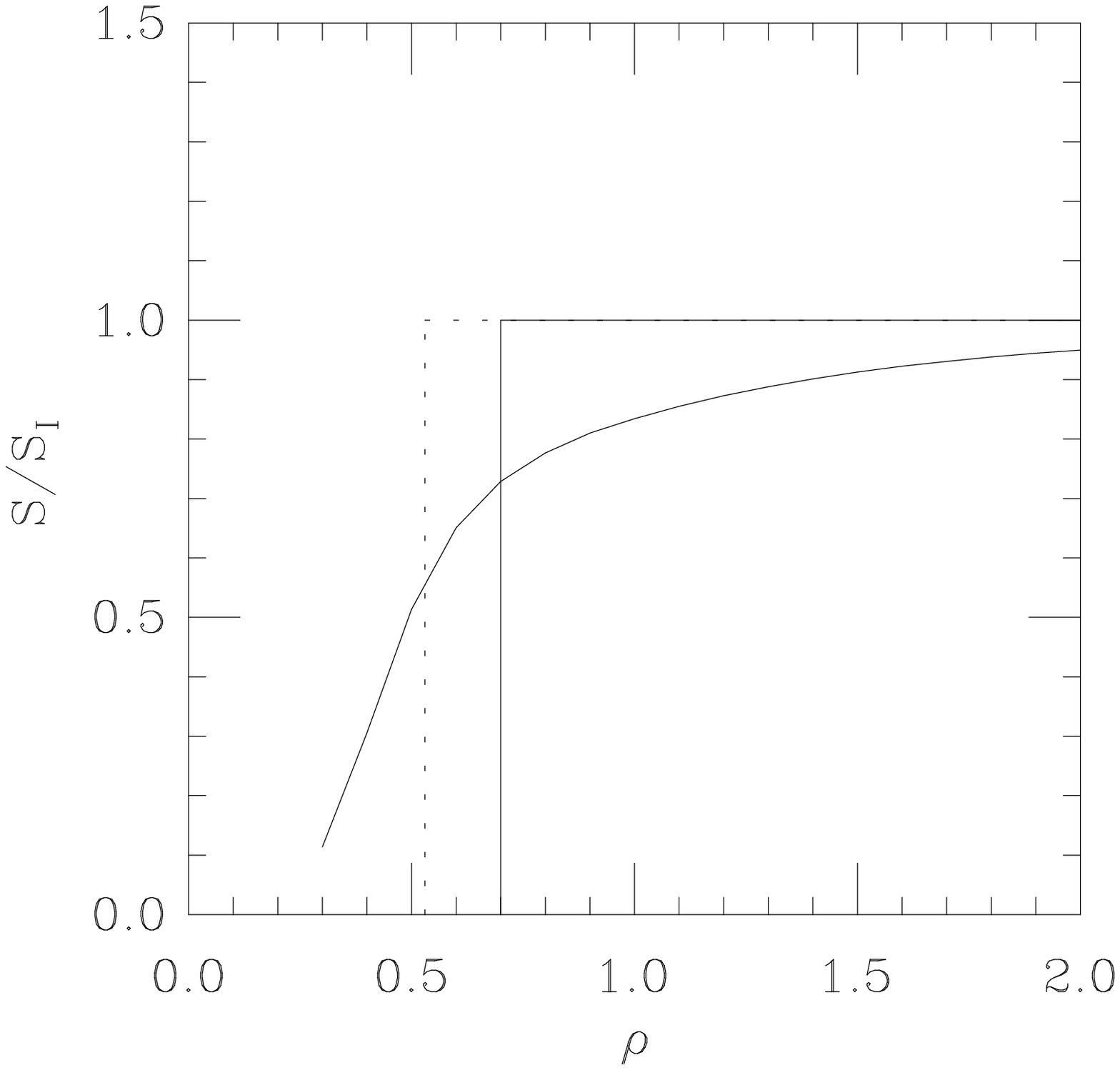}
\vskip 10mm
\end{center}
\caption{ Action of single blocked $c_2$
instanton configurations computed using the Wilson action,
displayed as in Fig. 5.}
\label{fig:sbc2w}
% FIGURE_FILE sbc2w.ax
\end{figure}

We feel it is important to demonstrate the (non-) scale invariant 
properties of the
Wilson action and the importance of inverse blocking in defining the topological
charge. In Fig.  \ref{fig:sbc1w}
-\ref{fig:sbc2w}
we show the action of our once  blocked $c_1$ and
$c_2$ instantons measured with the Wilson action. The dotted line indicates the
$Q=0 \to Q=1$ boundary on the original coarse lattice while the solid line is the
boundary on the inverse blocked configuration. The inverse blocking was done
with the FP action and RG transformation described in Sect. 2 and is not necessarily
correct for the Wilson action. Nevertheless, 
the difference between the two definitions is
obvious. For the $c_2$ once blocked instantons, for example, the boundary moves
from $\rho=0.54$ to $\rho=0.70$ -- enough to overcome the entropy problem
discussed in Sect. 2!

We want to emphasize one more time that our definition of the topological charge
has to be used with the corresponding FP action and it cannot be automatically
applied for the plaquette Wilson action.

\section{Conclusions}

In this paper we  have  presented a theoretically consistent definition 
of the topological charge  based on a renormalization group transformation 
and studied the scale invariant instanton solutions of
the corresponding FP action. 

We have demonstrated that the FP action shows the desired scale invariance, 
i.e., that instanton solutions with non-zero topological charge have
action $S(\rho) \ge 8 \pi^2/g^2$. 
For small radius
 smooth instanton configurations we showed that
the RG topological charge is different from the direct geometrical definition 
and only the former is consistent with the scale invariance of the FP
 action,  as  the
direct geometric method identifies dislocations with $S(\rho)<8\pi^2/g^2$  with
$Q=1$ instantons.

Our definition of the topological charge works only with 
the FP action of a given
RG transformation and cannot be used with 
arbitrary, non-FP actions. In the present
work we have studied only one renormalization group 
transformation but the method
generalizes easily and  the qualitative features should be the same
for any RGT.

The extension of the techniques we have described
to SU(3) gauge theory is straightforward. The only complication
will be the necessity to use the SU(3) analog of the geometric
definition for topological charge\cite{TOPOSU3}.

A measurement of the topological susceptibility using an eight
parameter approximate FP action valid for small correlation lengths 
 is  the subject of another paper\cite{INSTANTON2}.

\section{Acknowledgements}
We would like to thank M. M\"uller-Preussker for providing us with
a copy of the DESY program for measuring topological charge. 
We very much want to thank
P.~Hasenfratz and F.~Niedermayer for
useful conversations.
We would like to thank  A. Barker,
M. Horanyi and the Colorado high energy experimental
group for allowing us to use their work stations. 
This work was supported by the U.S. Department of 
Energy and by the National Science Foundation.

\newcommand{\PL}[3]{{Phys. Lett.} {\bf #1} {(19#2)} #3}
\newcommand{\PR}[3]{{Phys. Rev.} {\bf #1} {(19#2)}  #3}
\newcommand{\NP}[3]{{Nucl. Phys.} {\bf #1} {(19#2)} #3}
\newcommand{\PRL}[3]{{Phys. Rev. Lett.} {\bf #1} {(19#2)} #3}
\newcommand{\PREPC}[3]{{Phys. Rep.} {\bf #1} {(19#2)}  #3}
\newcommand{\ZPHYS}[3]{{Z. Phys.} {\bf #1} {(19#2)} #3}
\newcommand{\ANN}[3]{{Ann. Phys. (N.Y.)} {\bf #1} {(19#2)} #3}
\newcommand{\HELV}[3]{{Helv. Phys. Acta} {\bf #1} {(19#2)} #3}
\newcommand{\NC}[3]{{Nuovo Cim.} {\bf #1} {(19#2)} #3}
\newcommand{\CMP}[3]{{Comm. Math. Phys.} {\bf #1} {(19#2)} #3}
\newcommand{\REVMP}[3]{{Rev. Mod. Phys.} {\bf #1} {(19#2)} #3}
\newcommand{\ADD}[3]{{\hspace{.1truecm}}{\bf #1} {(19#2)} #3}
\newcommand{\PA}[3] {{Physica} {\bf #1} {(19#2)} #3}
\newcommand{\JE}[3] {{JETP} {\bf #1} {(19#2)} #3}
\newcommand{\FS}[3] {{Nucl. Phys.} {\bf #1}{[FS#2]} {(19#2)} #3}

%\bibitem{gasser}
%R. Willey and H. Yu,
%\PR{D26}{82}{3287};\ADD{D27}{87}{1234}.
%\bibitem{jhgf}
%G. 't Hooft, \NP{B72}{79}{461};\\
%E. Witten, \NP{B156}{79}{269};
%\ADD{B160}{79}{57};

\end{document}